\documentclass[usenatbib]{mn2e} 
\usepackage{graphicx,fixltx2e,amsmath,amssymb,amsfonts,latexsym,wasysym}

\def\rpd{\hbox{rad\,d$^{-1}$}}

\def\chisqr{\hbox{$\chi^2_{\rm r}$}}
\def\msun{\hbox{${\rm M}_{\odot}$}}

\def\rsun{\hbox{${\rm R}_{\odot}$}}

\def\rstar{\hbox{$R_{\star}$}}

\def\teff{\hbox{$T_{\rm eff}$}}
\def\logg{\hbox{$\log g$}}
\def\sn{\hbox{S/N}}

\def\kms{\hbox{km\,s$^{-1}$}}
\def\vsini{\hbox{$v \sin i$}}

\def\V{\hbox{${\rm V}$}}
\def\BmV{\hbox{${\rm B-V}$}}
\def\VmRj{\hbox{${\rm V-R_{\rm J}}$}}
\def\AV{\hbox{$A_{\rm V}$}}

\def\degr{\hbox{$^\circ$}}

\def\omeq{\hbox{$\Omega_{\rm eq}$}}
\def\dom{\hbox{$d\Omega$}}
\def\Porb{\hbox{$P_{\rm orb}$}} 
\def\Prot{\hbox{$P_{\rm rot}$}}

\newcommand{\caii}{\hbox{Ca$\;${\sc ii}}}
\newcommand{\fei}{\hbox{Fe$\;${\sc i}}}

\newcommand{\hei}{\hbox{He$\;${\sc i}}}

\begin{document}

\title[Magnetometry \& velocimetry of the wTTS LkCa~4]{Modelling the magnetic activity \& filtering radial velocity curves of young Suns : the weak-line T~Tauri star LkCa~4}
\makeatletter

\def\newauthor{%
  \end{author@tabular}\par
  \begin{author@tabular}[t]{@{}l@{}}}
\makeatother
 
\author[J.-F.~Donati et al.]
{\vspace{1.7mm}
J.-F.~Donati$^1$\thanks{E-mail: 
jean-francois.donati@irap.omp.eu }, 
E.~H\'ebrard$^1$, G.~Hussain$^{2,1}$, C.~Moutou$^3$, K.~Grankin$^4$, \\ 
\vspace{1.7mm}
{\hspace{-1.5mm}\LARGE\rm
I.~Boisse$^5$, J.~Morin$^6$, S.G.~Gregory$^7$, A.A.~Vidotto$^8$, J.~Bouvier$^9$, S.H.P.~Alencar$^{10}$, } \\ 
\vspace{1.7mm}
{\hspace{-1.5mm}\LARGE\rm
X.~Delfosse$^9$, R.~Doyon$^{11}$, M.~Takami$^{12}$, M.M.~Jardine$^7$, R.~Fares$^7$, A.C.~Cameron$^7$, } \\ 
\vspace{1.7mm}
{\hspace{-1.5mm}\LARGE\rm
F.~M\'enard$^{13}$, C.~Dougados$^{13,9}$, G.~Herczeg$^{14}$ and the MaTYSSE collaboration} \\
$^1$ Universit\'e de Toulouse / CNRS-INSU, IRAP / UMR 5277, Toulouse, F--31400 France \\ 
$^2$ ESO, Karl-Schwarzschild-Str.\ 2, D-85748 Garching, Germany \\ 
$^3$ CFHT Corporation, 65-1238 Mamalahoa Hwy, Kamuela, Hawaii 96743, USA \\ 
$^4$ Crimean Astrophysical Observatory, Nauchny, Crimea 298409, Russia \\
$^5$ Universit\'e Aix-Marseille / CNRS-INSU, LAM / UMR~7326, 13388 Marseille, FRANCE \\ 
$^6$ Universit\'e Montpellier / CNRS-INSU, LUPM / UMR 5299, 34095 Montpellier, France \\
$^7$ School of Physics and Astronomy, Univ.\ of St~Andrews, St~Andrews, Scotland KY16 9SS, UK \\ 
$^8$ Observatoire de Gen\`eve, Chemin des Maillettes 51, CH-1290 Versoix, Suisse \\ 
$^9$ Universit\'e Joseph Fourier / CNRS-INSU, IPAG / UMR 5274, Grenoble, F-38041, France \\ 
$^{10}$ Departamento de F\`{\i}sica -- ICEx -- UFMG, Av. Ant\^onio Carlos, 6627, 30270-901 Belo Horizonte, MG, Brazil \\ 
$^{11}$ D\'epartement de physique, Universit\'e de Montr'eal, C.P.~6128, Succursale Centre-Ville, Montr\'eal, QC, Canada  H3C 3J7 \\ 
$^{12}$ Institute of Astronomy and Astrophysics, Academia Sinica, PO Box 23-141, 106, Taipei, Taiwan \\ 
$^{13}$ UMI-FCA, CNRS/INSU, France (UMI 3386), and Universidad de Chile, Santiago, Chile  \\ 
$^{14}$ Kavli Institute for Astronomy and Astrophysics, Peking University, Yi He Yuan Lu 5, Haidian Qu, Beijing 100871, China  
}

\date{2014 August, MNRAS in press}
\maketitle
 
\begin{abstract}  
We report results of a spectropolarimetric and photometric monitoring of the weak-line T~Tauri star LkCa~4 
within the MaTYSSE programme, involving ESPaDOnS at the Canada-France-Hawaii Telescope.
Despite an age of only 2~Myr and a similarity with prototypical classical T~Tauri stars, LkCa~4 shows 
no evidence for accretion and probes an interesting transition stage for star and planet formation.  
Large profile distortions and Zeeman signatures are detected in the unpolarized and 
circularly-polarized lines of LkCa~4 using Least-Squares Deconvolution (LSD), indicating the 
presence of brightness inhomogeneities and magnetic fields at the surface of LkCa~4.  

Using tomographic imaging, we reconstruct brightness and magnetic maps of LkCa~4 from sets 
of unpolarized and circularly-polarized LSD profiles.  The large-scale field is strong and mainly 
axisymmetric, featuring a $\simeq$2~kG poloidal component and a $\simeq$1~kG toroidal component 
encircling the star at equatorial latitudes - the latter making LkCa~4 markedly different from 
classical T~Tauri stars of similar mass and age.  The brightness map includes a dark spot overlapping the 
magnetic pole and a bright region at mid latitudes - providing a good match to the contemporaneous 
photometry.  We also find that differential rotation at the surface of LkCa~4 is small, typically $\simeq$5.5$\times$ 
weaker than that of the Sun, and compatible with solid-body rotation.  

Using our tomographic modelling, we are able to filter out the activity jitter in the RV curve of LkCa~4 
(of full amplitude 4.3~\kms) down to a rms precision of 0.055~\kms.  Looking for hot Jupiters around young Sun-like 
stars thus appears feasible, even though we find no evidence for such planets around LkCa~4.  
\end{abstract}

\begin{keywords} 
stars: magnetic fields --  
stars: formation -- 
stars: imaging -- 
stars: rotation -- 
stars: individual:  LkCa~4 --
techniques: polarimetric
\end{keywords}

\section{Introduction} 
\label{sec:int}

Magnetic fields are known to have a significant impact at early stages of evolution, when stars and their planets form 
from collapsing dense pre-stellar cores, progressively flattening into large-scale magnetized accretion discs and 
finally settling as pre-main-sequence (PMS) stars surrounded by protoplanetary discs \citep[e.g.,][]{Andre09, Donati09}.  
At an age of 1-10~Myr, low-mass PMS stars have emerged from their dust cocoons and are still in a phase of gravitational 
contraction towards the main sequence (MS).  They are either classical T-Tauri stars (cTTSs) when still surrounded by a 
massive (presumably planet-forming) accretion disc or weak-line T-Tauri stars (wTTSs) when their discs have mostly dissipated.  
Both cTTSs and wTTSs have been the subject of intense scrutiny at all wavelengths in recent decades given their interest 
for benchmarking the scenarios currently invoked to explain low-mass star and planet formation.  

Magnetic fields of cTTSs play a key role in controlling accretion processes and in triggering outflows, and thus largely dictate the 
angular momentum evolution of low-mass PMS stars \citep[e.g.,][]{Bouvier07, Frank14}.  More specifically, large-scale fields of cTTSs 
can evacuate the central regions of accretion discs (where dust has already sublimated and only gas is left), 
funnel the disc material onto the stars, and enforce corotation between 
cTTSs and their inner-disc Keplerian flows, thus causing cTTSs to rotate much more slowly than expected from the contraction 
and accretion of the high-angular-momentum disc material.  Although magnetic fields of cTTSs were first detected about 15 years 
ago \citep[e.g.,][]{Johns99b, Johns07}, their topologies remained elusive for a long time.  In particular, the large-scale 
fields of cTTSs have only recently been unveiled for a dozen cTTSs \citep[e.g.,][]{Donati07, Hussain09, Donati10b, Donati13}, 
largely thanks to the MaPP (Magnetic Protostars and Planets) Large Observing Programme allocated on the 3.6~m 
Canada-France-Hawaii Telescope (CFHT) with the ESPaDOnS high-resolution spectropolarimeter, over a timespan of 9 semesters 
(2008b-2012b, 550~hr of clear time out of an allocation of 690~hr).  This first survey revealed that the large-scale 
fields of cTTSs depend mostly on the internal structure of 
the PMS star.  More specifically, these large-scale fields, although often more complex than pure dipoles and featuring a 
significant (sometimes dominant) octupolar component, remain rather simple when the PMS star is still fully or largely 
convective, but become much more complex when the PMS star turns mostly radiative \citep{Gregory12, Donati13}.  This survey also 
showed that these fields are likely of dynamo origin, being variable on timescales of years \citep[e.g.,][]{Donati11, 
Donati12, Donati13} and similar to those of mature stars with comparable internal structures \citep{Morin08b}.  

Comparatively little is known about the large-scale magnetic topologies of wTTSs, with only 1 such PMS star imaged to date 
\citep[namely V410~Tau, ][]{Skelly10}.  Yet, being the missing link between cTTSs and MS low-mass stars, wTTSs are key targets 
to study the magnetic topologies and associated winds with which disc-less PMS stars initiate their unleashed spin-up as they 
contract towards the MS. The results obtained for V410~Tau, showing a complex field with a significant toroidal and a 
non-axisymmetric poloidal component despite being fully convective, are in surprising contrast with those for largely- and 
fully-convective cTTSs and mature M dwarfs, harbouring rather simple poloidal fields \citep{Donati13, Morin08b}.  Observing a significant 
number of wTTSs in a few of the nearest star forming regions is needed to have a better view of their magnetic topologies.  
It will also allow us to study in a quantitative way the magnetic winds of wTTSs and the corresponding spin-down 
rates \citep{Vidotto14}.  Last but not least, it should give the opportunity to filter-out most of the activity jitter from radial 
velocity (RV) curves of wTTSs (using spectropolarimetry as the most reliable proxy of surface activity) for potentially detecting 
hot Jupiters (hJs) around wTTSs.  This is what the new MaTYSSE (Magnetic Topologies of Young Stars and the Survival of close-in 
giant Exoplanets) Large Programme, recently allocated at CFHT over a timescale of 8 semesters (2013a-2016b, 510~hr) 
with complementary observations with the NARVAL spectropolarimeter on the 2-m T\'elescope Bernard Lyot at Pic du Midi in France and 
with the HARPS spectropolarimeter at the 3.6-m ESO Telescope at La Silla in Chile, is aiming at.  Although mainly focussed on 
studying wTTSs, MaTYSSE also includes a long-term monitoring of a few cTTSs to better estimate the amount by which their 
large-scale magnetic topologies (and therefore their central magnetospheric cavities) vary with time and the implications of such 
fluctuations on the survival of potential hJs.  

We present in this paper phase-resolved spectropolarimetric observations of the wTTSs LkCa~4 collected in the framework 
of MaTYSSE with ESPaDOnS at the CFHT in 2014 January, complemented with contemporaneous photometric observations secured at 
the 1.25-m telescope of the Crimean Astrophysical Observatory (CrAO).  After documenting the observations (Sec.~\ref{sec:obs}) 
and briefly recalling the spectral characteristics and corresponding evolutionary status of LkCa~4 (Sec.~\ref{sec:lk}), 
we describe the results obtained by applying our tomographic modelling technique to this new data set 
(Sec.~\ref{sec:mod});  we also outline its practical use for filtering out the activity jitter from RV curves of young Sun-like 
stars and for detecting hJs potentially orbiting around them (Sec.~\ref{sec:fil}).  We finally summarize the main outcome of our 
study, discuss its implications for our understanding of low-mass star and planet formation, and conclude with prospects and 
long-term plans (Sec.~\ref{sec:dis}).

\section{Observations}
\label{sec:obs}

Spectropolarimetric observations of LkCa~4 were collected in 2014 January, using the high-resolution
spectropolarimeter ESPaDOnS at the 3.6-m Canada-France-Hawaii Telescope (CFHT) atop Mauna Kea (Hawaii).  
ESPaDOnS collects stellar spectra spanning the entire optical domain (from 370 to 1,000~nm) at a resolving power of 65,000 
(i.e., resolved velocity element of 4.6~\kms) over the full wavelength, in either circular or linear polarisation \citep{Donati03}.   
A total of 12 circularly-polarized (Stokes $V$) and unpolarized (Stokes $I$) spectra were collected over a timespan of 13 nights,
corresponding to about 4 rotation cycles of LkCa~4;  the time sampling is regular except for a gap of 3~nights due to bad weather 
during the second rotation cycle (Jan 12--14).  

All polarisation spectra consist of 4 individual subexposures (each lasting 946~s) taken in different
polarimeter configurations to allow the removal of all spurious polarisation signatures at first order.
All raw frames are processed as described in the previous papers of the series
\citep[e.g.,][]{Donati10b, Donati11}, to which the reader is referred for more information.
The peak signal-to-noise ratios (\sn, per 2.6~\kms\ velocity bin) achieved on the
collected spectra range between 120 and 180 (with a median of 160),
depending mostly on weather/seeing conditions.
The full journal of observations is presented in Table~\ref{tab:log}.

\begin{table}
\caption[]{Journal of ESPaDOnS observations of LkCa~4 collected in 2014~January.
Each observation consists of a sequence of 4 subexposures, lasting 946~s each.
Columns $1-4$ respectively list (i) the UT date of the observation, (ii) the 
corresponding UT time (at mid-exposure), (iii) the Barycentric Julian Date (BJD) in 
excess of 2,456,000, and (iv) the peak signal to noise ratio (per 2.6~\kms\ velocity bin) 
of each observation.  Column 5 lists the rms noise level (relative to the unpolarized continuum level
$I_{\rm c}$ and per 1.8~\kms\ velocity bin) in the circular polarization profile
produced by Least-Squares Deconvolution (LSD), while column~6 indicates the
rotational cycle associated with each exposure (using the ephemeris given by
Eq.~\ref{eq:eph}).  }
\begin{tabular}{cccccc}
\hline
Date   & UT      & BJD      & \sn\ & $\sigma_{\rm LSD}$ & Cycle \\
(2014) & (h:m:s) & (2,456,000+) &      &   (0.01\%)  &   \\
\hline
Jan 08 & 05:11:16 & 665.72044 & 120 & 4.5 & 0.006 \\
Jan 09 & 08:18:38 & 666.85048 & 170 & 3.1 & 0.341 \\
Jan 10 & 06:26:43 & 667.77270 & 180 & 3.3 & 0.614 \\
Jan 11 & 08:46:49 & 668.86991 & 160 & 3.5 & 0.939 \\
Jan 15 & 09:29:48 & 672.89947 & 120 & 4.5 & 2.134 \\
Jan 16 & 08:05:24 & 673.84079 & 140 & 4.1 & 2.413 \\
Jan 17 & 06:30:10 & 674.77460 & 160 & 3.6 & 2.690 \\
Jan 18 & 05:39:50 & 675.73957 & 160 & 3.5 & 2.976 \\
Jan 19 & 07:00:21 & 676.79539 & 180 & 3.0 & 3.288 \\
Jan 20 & 08:47:47 & 677.86992 & 160 & 3.5 & 3.607 \\
Jan 21 & 05:43:35 & 678.74193 & 170 & 3.3 & 3.865 \\
Jan 21 & 09:24:03 & 678.89502 & 180 & 3.2 & 3.911 \\
\hline
\end{tabular}
\label{tab:log}
\end{table}

Rotational cycles of LkCa~4 (noted $E$ in the following equation) are computed from Barycentric Julian Dates (BJDs)
according to the ephemeris:
\begin{equation}
\mbox{BJD} \hbox{\rm ~(d)} = 2456665.7 + 3.374 E
\label{eq:eph}
\end{equation}
in which the photometrically-determined rotation period \Prot\ \citep[equal to 3.374~d to an accuracy better than 0.01~d, e.g.,][]{Grankin08} is taken from the literature 
and the initial Julian date (2456665.7~d) is chosen arbitrarily.  

Least-Squares Deconvolution \citep[LSD,][]{Donati97b} was applied to all observations.   
The line list we employed for LSD is computed from an {\sc Atlas9} LTE model atmosphere \citep{Kurucz93} 
featuring $\teff=4,250$~K and $\logg=4.0$, appropriate for LkCa~4 (see Sec.~\ref{sec:lk}).
As usual, only moderate to strong atomic spectral lines are included in this list \citep[see, e.g.,][for more details]{Donati10b}.
Altogether, about 7,800 spectral features (with about 40\% from \fei) are used in this process.
Expressed in units of the unpolarized continuum level $I_{\rm c}$, the average noise levels of the resulting Stokes $V$ 
LSD signatures range from 3.0 to 4.5$\times10^{-4}$ per 1.8~\kms\ velocity bin (median value 3.5$\times10^{-4}$).  

Zeeman signatures are detected at all times in Stokes $V$ LSD profiles (see Fig.~\ref{fig:lsd} for an example), 
featuring large amplitudes of typically 1\%, i.e., already indicative of strong large-scale fields.  
Significant distortions are also visible most of the time 
in Stokes $I$ LSD profiles, strongly suggesting the presence of brightness inhomogeneities covering 
a large fraction of the surface of LkCa~4 at the time of our observations.  

Contemporaneous BVR$_{\rm J}$ photometric observations were also collected from the CrAO 1.25~m telescope between 2013 Sep and 
Dec, indicating that LkCa~4 was undergoing brightness modulations \citep[with a period compatible with the rotation period of][]{Grankin08} of full 
amplitude 0.35~mag in \V\ (see Table~\ref{tab:pho}) - on the low side of what is reported for this star \citep[with amplitudes 
in \V\ fluctuations ranging from 0.35 to 0.79,][]{Grankin08}.  Again, this is clearly indicative of large spots at the surface of LkCa~4.  

\begin{table}
\caption[]{Journal of contemporaneous CrAO multicolour photometric observations of LkCa~4 collected in late 2013, 
respectively listing the UT date and Heliocentric Julian Date (HJD) of the observation, the measured \V\ magnitude, 
\BmV\ and \VmRj\ Johnson photometric colours, and the corresponding rotational phase (using again the ephemeris given by
Eq.~\ref{eq:eph}). } 
\begin{tabular}{cccccc}
\hline
Date   & HJD      & \V    & \BmV & \VmRj & Phase \\
(2013) & (2,456,000+) & (mag) &      &      &       \\
\hline
Sep 01 & 537.5417 & 12.780 & 1.469 & 1.431 & 0.016 \\ 
Sep 11 & 547.5133 & 12.819 & 1.496 & 1.421 & 0.971 \\ 
Oct 11 & 577.5613 & 12.942 & 1.507 & 1.460 & 0.877 \\ 
Oct 27 & 593.5458 & 12.821 & 1.351 & 1.440 & 0.615 \\ 
Oct 29 & 595.5525 & 12.699 & 1.409 & 1.420 & 0.209 \\ 
Oct 30 & 596.5834 & 12.669 & 1.419 & 1.406 & 0.515 \\ 
Nov 08 & 605.5484 & 12.749 & 1.476 & 1.421 & 0.172 \\ 
Nov 09 & 606.5891 & 12.658 & 1.447 & 1.399 & 0.480 \\ 
Nov 10 & 607.4750 & 13.010 & 1.504 & 1.448 & 0.743 \\ 
Dec 03 & 630.4152 & 12.730 & 1.478 & 1.432 & 0.542 \\ 
\hline
\end{tabular}
\label{tab:pho}
\end{table}

\begin{figure}
\includegraphics[scale=0.35,angle=-90]{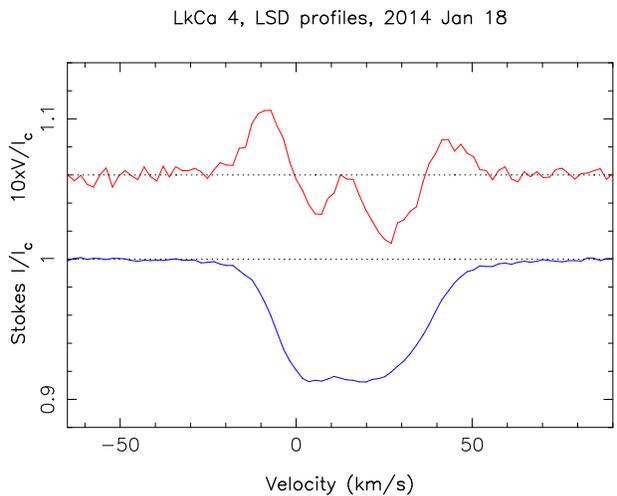}
\caption[]{LSD circularly-polarized (Stokes $V$) and unpolarized (Stokes $I$)
profiles of LkCa~4 (top/red, bottom/blue curves respectively) collected on
2014~Jan.~18 (cycle 2.976).  A clear and complex Zeeman signature (with a full amplitude of 1\%)
is detected in the LSD Stokes $V$ profile, in conjunction with the unpolarised line profile.
The mean polarization profile is expanded by a factor of 10 and shifted upwards
by 1.06 for display purposes.  }
\label{fig:lsd}
\end{figure}

\section{Evolutionary status of LkCa~4}
\label{sec:lk}

LkCa~4 is a single star, showing no spectroscopic nor imaging evidence for a companion \citep[e.g.,][]{White01, Kraus11}.  
Applying the automatic spectral classification tool especially developed in the context of MaPP and MaTYSSE, 
inspired from that of \citet{Valenti05} and discussed in a previous paper \citep{Donati12}, we find that 
the photospheric temperature and logarithmic gravity of LkCa~4 are respectively equal to 
$\teff=4,100\pm50$~K and $\logg=3.8\pm0.1$ (with $g$ in cgs units).  This temperature estimate agrees well with that of 
\citet{Grankin13} derived from photometric colors, but is significantly larger than that obtained by \citet{Herczeg14} from fits to flux 
calibrated low-resolution spectra\footnote{We suspect that this discrepancy is partly attributable to the different visual extinction they 
derived, and partly to the different spectral proxies used to estimate temperature (molecular features vs atomic lines in our case) rendering 
their technique presumably more sensitive to the presence of large cool spots at the surface of the star}.  In addition to being based 
on higher-resolution spectra and to providing a better match to photometry, our temperature estimate also has the advantage of 
being homogeneous with those derived in our previous studies of cTTSs \citep[e.g.,][]{Donati13}, rendering internal comparisons simpler and more meaningful.  
From the \BmV\ index expected at this temperature \citep[equal to $1.20\pm0.02$,][]{Pecaut13} and the averaged value measured for LkCa~4 
\citep[equal to $\simeq$1.42, see][, see also Table~\ref{tab:pho}]{Grankin08}, we find that the amount of visual extinction 
\AV\ that LkCa~4 is suffering is moderate, equal to $0.68\pm0.15$.  Similarly, the visual bolometric correction 
expected for this temperature is found to be $-1.00\pm0.05$ \citep{Pecaut13}.  

Accurate distances to young stars near LkCa~4 in Taurus measured with Very Long Base Interferometry \citep[namely the wTTSs Hubble~4, HDE~283572 and 
the quadruple wTTSs system V773~Tau, ][]{Torres07, Torres12} are all consistent with an average distance of 130~pc to within better than 5~pc, 
implying a distance modulus of $5.57\pm0.08$ for LkCa~4 (provided it belongs to the same star-formation complex).  
Assuming for LkCa~4 an unspotted \V\ magnitude of $12.1\pm0.2$ \citep[given the maximum \V\ magnitude of 12.3 reported 
by][and taking into account a spot coverage of $\simeq$20\% at maximum brightness, typical for active stars, see also Sec.~\ref{sec:mod}]{Grankin08}, 
we finally obtain a bolometric magnitude of $4.85\pm0.27$, or equivalently a logarithmic luminosity 
(relative to the Sun) of $-0.04\pm0.11$.  

\begin{figure}
\includegraphics[scale=0.35,angle=-90]{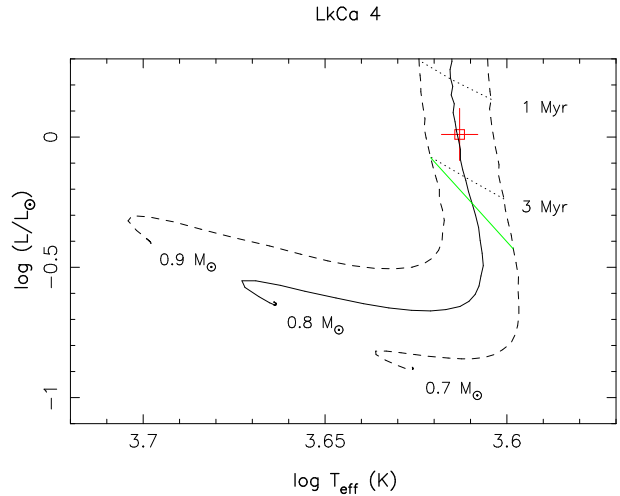}
\caption[]{Observed (open square and error bars) location of LkCa~4 in the HR diagram.
The PMS evolutionary tracks and corresponding isochrones \citep{Siess00} assume solar
metallicity and include convective overshooting.  The green line depicts where models
predict PMS stars start developing their radiative core as they contract towards the main
sequence.  }
\label{fig:hrd}
\end{figure}

Considering that the rotation period of LkCa~4 is 3.374~d \citep{Grankin08} and that its line-of-sight-projected equatorial 
rotation velocity \vsini\ is $28.0\pm0.5$~\kms\ (see Sec.~\ref{sec:mod}), we can infer that $\rstar \sin i = 1.87\pm0.03$~\rsun\ 
where \rstar\ and $i$ respectively note the radius of LkCa~4 and the inclination of its rotation axis to the line of sight.  
Given our photospheric temperature estimate of $\teff=4,100\pm50$~K, we derive a minimum logarithmic luminosity (with respect to 
that of the Sun) of $-0.05\pm0.04$.  This clearly indicates that $i$ is large, typically ranging from $\simeq$60\degr\ (for our 
upper limit of $\simeq$0.07 for the logarithmic luminosity) up to $\simeq$90\degr\ (for the minimum logarithmic luminosity of $-0.05$).  

In the forthcoming study, we adopt for LkCa~4 an intermediate inclination of $i=70$\degr, corresponding to a relative logarithmic 
luminosity of $0.0\pm0.1$ and  a radius of $2.0\pm0.2$~\rsun.  
Using the evolutionary models of \citet{Siess00} (assuming solar metallicity and including convective overshooting), 
we obtain that LkCa~4 is a $0.79\pm0.05$~\msun\ star of age $\simeq$2~Myr, indicating that this star is still fully convective 
and structurally similar to the prototypical cTTSs AA~Tau and BP~Tau.  Our estimates are in reasonable agreement with those found in 
the refereed literature \citep[e.g.,][]{Grankin13}.  

We finally report that \caii\ IRT lines of LkCa~4 feature core emission, with an average equivalent width of the emission core equal to 
$\simeq$9~\kms, i.e.\ very close to the amount expected from chromospheric emission for such PMS stars \citep[e.g.,][]{Donati13};  similarly, the \hei\ $D_3$ 
line is barely visible (average equivalent width of $\simeq$10~\kms).  In both cases, this is a reliable indication that LkCa~4 is 
no longer accreting material on its surface, thus further confirming its wTTS status in agreement with other recent studies \citep[e.g.,][]{Gomez13}.  
This makes LkCa~4 a very interesting target for probing a transition stage that is critical for our understanding of star and planet formation.

\section{Tomographic modelling}
\label{sec:mod}

In this section, we describe the application of our dedicated stellar-surface tomographic-imaging tool to our new spectropolarimetric data set.  
The software package we are using for this purpose is based on the principles of maximum-entropy image reconstruction, assuming that the 
observed variability is mainly caused by rotational modulation (with an added option for differential rotation).  Since the initial release 
\citep{Brown91, Donati97c}, this package underwent several significant upgrades \citep[e.g.,][]{Donati01, Donati06b}, the most recent one 
being its re-profiling to the specific needs of MaPP observations \citep{Donati10b}.  The reader is referred to these papers for more general 
details on the imaging method.  

We use here this last version in a slightly modified configuration that allows a more accurate modelling of the MaTYSSE data.  More specifically, 
the code is set up to invert (both automatically and simultaneously) time series of Stokes $I$ and $V$ LSD profiles into brightness and magnetic 
maps of the stellar surface.  The reconstructed brightness map is now allowed to include not only cool regions (as before) but also bright plages 
as well - known to be participating to the activity of very active stars, especially those like LkCa~4 featuring extreme levels of photometric 
variability.  This is achieved by allowing the reconstructed surface brightness (normalised to that of the photosphere) to vary both 
below and above 1, rather than being constrained to a ]0,1] interval, and by modifying accordingly the entropy associated to this image quantity.  

While using sets of Stokes $I$, $V$, $Q$ and $U$ (rather than only $I$ and $V$) profiles could in principle be useful to further constrain the imaging 
process and eliminate potential imaging artifacts \citep[e.g.,][]{Donati01, Kochukhov02}, it turns out to be unfeasible in practice for stars as faint as TTSs.  
Stokes $Q$ and $U$ Zeeman signatures are indeed much weaker than their Stokes $V$ equivalents;  moreover, they depend in a much stronger way on the actual 
Zeeman patterns of individual lines, making them far less suited to LSD-like multiline techniques without which they are basically impossible to detect.  
As a result and for a given amount of observing time, tomographic magnetic imaging of TTSs ends up being more efficient when applied to data sets limited to 
Stokes $I$ and $V$ profiles only, but featuring a phase coverage $3\times$ denser than if all Stokes profiles were recorded.  Moreover, by making adequate 
prior assumptions on the magnetic topologies to be imaged (described as a low-order multipolar expansion), one can largely suppress most residual imaging 
artifacts even when only Stokes $I$ and $V$ profiles are used \citep[e.g.][]{Donati01}.  

The local Stokes $I$ and $V$ profiles used to compute the disc-integrated average photospheric LSD profile of LkCa~4 are synthesized using 
Unno-Rachkovsky's analytical solution to the equations of polarized radiative transfer in a Milne-Eddington model atmosphere, known to provide a 
reliable description (including magneto-optical effects) of how shapes of line profiles are distorted in the presence of magnetic fields \citep[e.g.,][]{Landi04}.  
The main parameters of this local profile are similar to those used in our previous studies, the wavelength, Doppler width, equivalent width and 
Land\'e factor being respectively set to 670~nm, 1.8~\kms, 3.9~\kms\ and 1.2.  

As part of the imaging process, we obtain accurate estimates for several parameters of LkCa~4.  We find in particular that the average RV of LkCa~4 
is $16.8\pm0.1$~\kms\ (more about this in Sec.~\ref{sec:fil}) and that the \vsini\ is equal to $28\pm0.5$~\kms.  

\subsection{Brightness and magnetic imaging}

We show in Fig.~\ref{fig:fit} our set of Stokes $I$ and $V$ LSD profiles of LkCa~4 along with our fit to the data, and in Fig.~\ref{fig:map} 
the corresponding brightness and magnetic maps that we reconstruct from fitting these data.  
The fit we obtain corresponds to a reduced chi-square equal to 1, starting from an initial value of over 50 (for a null magnetic 
field and an unspotted brightness map), which further stresses the quality of our data set and the high performance of our imaging code 
at modeling the observed modulation of LSD profiles (also obvious from Fig.~\ref{fig:fit}).  Given this, the large \vsini\ and the reasonably dense phase 
coverage, we can safely claim that no significant imaging artifacts nor biases are expected in the reconstructed maps.

\begin{figure}
\center{
\hbox{\hspace{-2mm}
\includegraphics[scale=0.6,angle=-90]{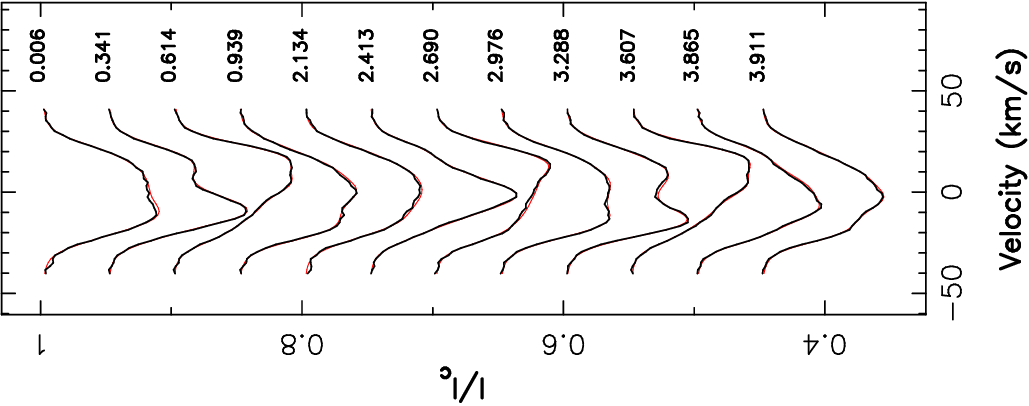}\hspace{3mm}
\includegraphics[scale=0.6,angle=-90]{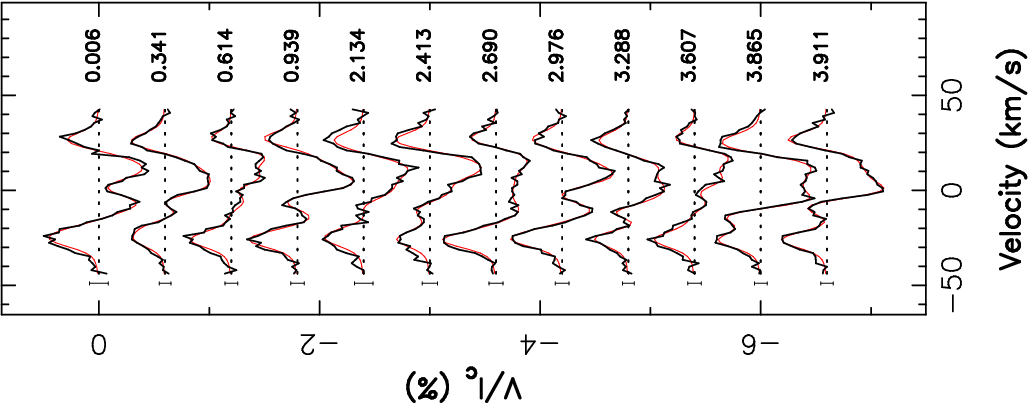}}}
\caption[]{Maximum-entropy fit (thin red line) to the observed (thick black line) Stokes $I$ (right panel) 
and Stokes $V$ (left panel) LSD photospheric profiles of LkCa~4 in 2014~Jan.  
Rotational cycles and 3$\sigma$ error bars (for Stokes $V$ profiles) are also shown next to each profile. 
This figure is best viewed in color.  }
\label{fig:fit}
\end{figure}

\begin{figure*}
\includegraphics[scale=0.6]{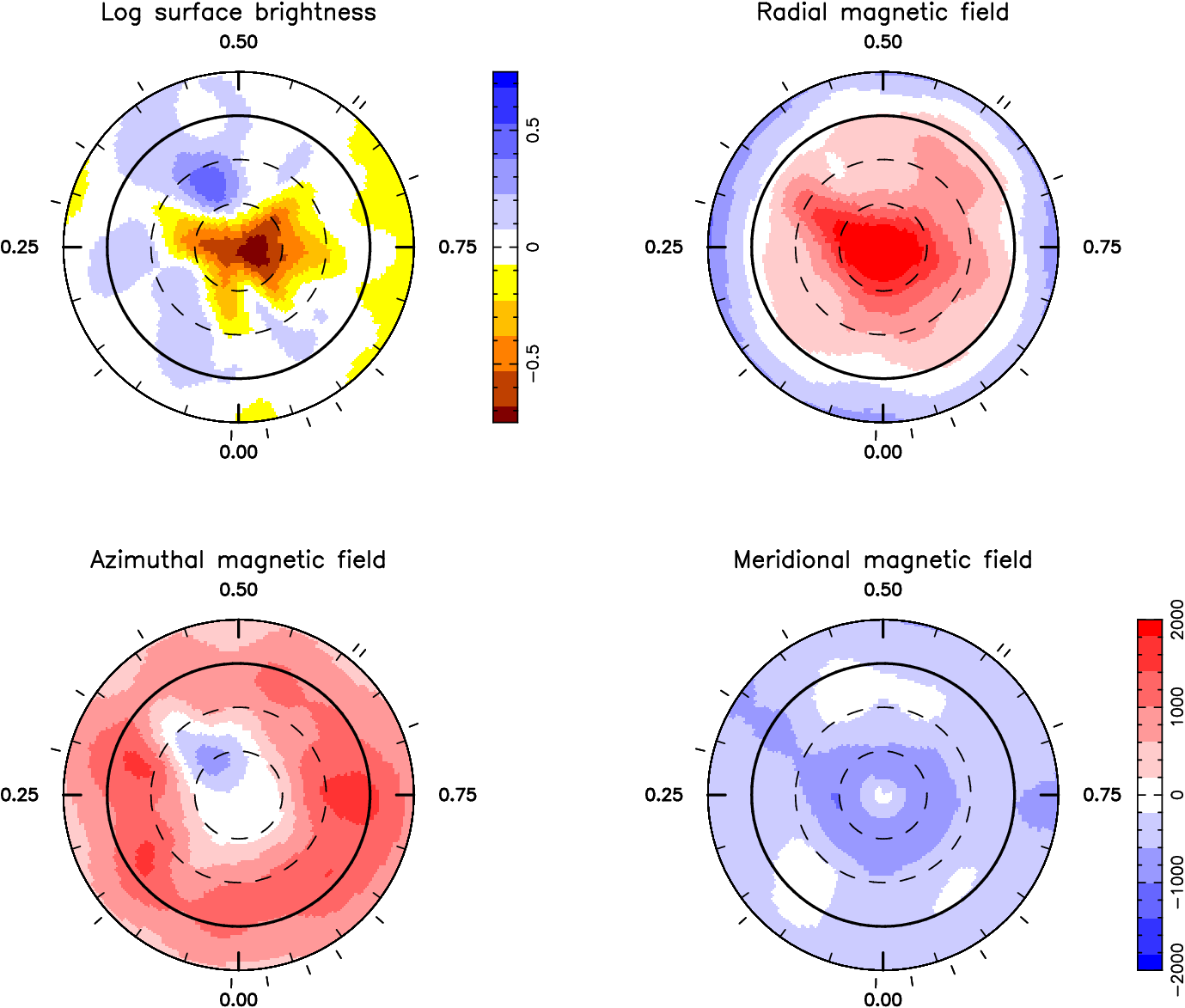}
\caption[]{Maps of the logarithmic brightness (relative to the quiet photosphere, top left panel), 
radial (top right), azimuthal (bottom left) and meridional (bottom right) components of the 
magnetic field $\bf B$ at the surface of LkCa~4 in 2014~Jan.  Magnetic fluxes are labelled in G.  
In all panels, the star is shown in flattened polar projection down to latitudes of $-30\degr$,
with the equator depicted as a bold circle and parallels as dashed circles.  Radial ticks around
each plot indicate phases of observations.  This figure is best viewed in color. }
\label{fig:map}
\end{figure*}

The main feature of the reconstructed brightness map (see top left panel of Fig.~\ref{fig:map}) is a cool spot of complex shape near the visible pole, 
which directly reflects that the mean Stokes $I$ LSD profile (not shown here, but resembling that at cycle 2.976, see Fig.~\ref{fig:lsd} and left panel of 
Fig.~\ref{fig:fit}) 
features a flatter bottom than a pure rotation profile;  although less dramatic for equator-on stars (with $i\geq60$\degr) than for pole-on ones ($i\leq30$\degr), 
this effect is nonetheless clearly detected in LkCa~4.  This polar spot is found to be highly non-axisymmetric and rather complex in shape, with as much as 3 
(and possibly up to 5) different appendages towards lower latitudes, as evidenced by the large level of temporal variability in the unpolarized LSD profiles.  
The second most prominent feature of the brightness map is the warm plage reconstructed at phase 0.42 and at mid latitudes, in which the local brightness 
rises up to $\simeq$50\% above that of the mean quiet photosphere.  Again, this directly reflects the very distorted shapes of Stokes $I$ LSD profiles at 
phases in the range 0.25--0.60, i.e., when the plage is most visible to the observer and generates conspicuous absorption dips in spectral lines (at cycles, 
e.g., 3.288, 0.341, 2.413, see left panel of Fig.~\ref{fig:fit}).  No similar fit to the data can be achieved if we force the code to reconstruct 
dark spots only at the surface of the star - an option implemented in our code and often used in Doppler imaging of cool stars to further constrain 
the inversion process with prior information on the image to be reconstructed.  We therefore conclude that bright plages are clearly present at the 
surface of LkCa~4, rendering this imaging option mandatory for modelling MaTYSSE data.  These two main surface features are found to cover altogether 
$\simeq$25\% of the overall stellar surface.  We stress that the photometric variations predicted by the 
reconstructed brightness maps are in very good agreement with the observations collected at CrAO (see Fig.~\ref{fig:pho}), even though they were not 
recorded simultaneously with (but rather 1--4 months before) our spectropolarimetric data set (see Sec.~\ref{sec:obs});  this suggests that the lifetime 
of brightness features at the surface of LkCa~4 is longer than a few months, fully compatible with previous conclusions from long-term photometric monitoring 
showing that light curves of LkCa~4 are remarkably stable in both shape and phase on timescales of several years \citep[see Fig.~5 in][]{Grankin08}.  

The magnetic map shows a rather simple large-scale magnetic structure featuring two main components.  The first one is a mostly-axisymmetric poloidal 
field enclosing $\simeq$70\% of the reconstructed magnetic energy;  94\% of the poloidal field energy gathers in spherical harmonics (SH) modes with 
$m<\ell/2$ ($\ell$ and $m$ denoting respectively the degrees and orders of the modes) while 86\% of it concentrates in the aligned dipole ($\ell=1$ and $m=0$) mode.  
This poloidal component can be approximated with a dominant 1.6~kG dipole tilted at $\simeq$10\degr\ to the line of sight (towards phase 0.75), 
with the addition of a 4$\times$ weaker ($\simeq$0.4~kG) mostly-axisymmetric octupolar component (roughly aligned with the dipole component), generating 
an intense radial field feature in 
excess of 2~kG at the visible pole of the star (see top right panel of Fig.~\ref{fig:map}) that coincides with the cool spot reconstructed 
in the brightness image.  We also note that the elongated shape of this magnetic pole is similar to that of the cool polar spot.  

The second main component of this large-scale field is a very significant, almost axisymmetric toroidal component showing up as an azimuthal field ring 
of $\simeq$1~kG encircling the star at equatorial latitudes (see bottom left panel of Fig.~\ref{fig:map}).  This feature directly reflects the nearly 
symmetric shape of the Stokes $V$ LSD profiles (about the centre of the line profile) at most rotation phases (see right panel of Fig.~\ref{fig:fit}), 
known to be a clear signature of a 
prominent azimuthal field ring at the surface of the star.  We also note the detection of an azimuthal field feature of opposite (i.e. clockwise) 
polarity in the immediate vicinity of (and possibly coincident with) the warm plage detected in the brightness image.  
SH expansions describing the reconstructed field presented in Fig.~\ref{fig:map} are limited to terms with $\ell\leq10$; 
only marginal changes to the solution are observed when increasing the maximum $\ell$ value from 10 to 15, demonstrating that most of the signal 
detected in Stokes $V$ LSD profiles of LkCa~4 concentrates at large spatial scales.  

\begin{figure}
\includegraphics[scale=0.35,angle=-90]{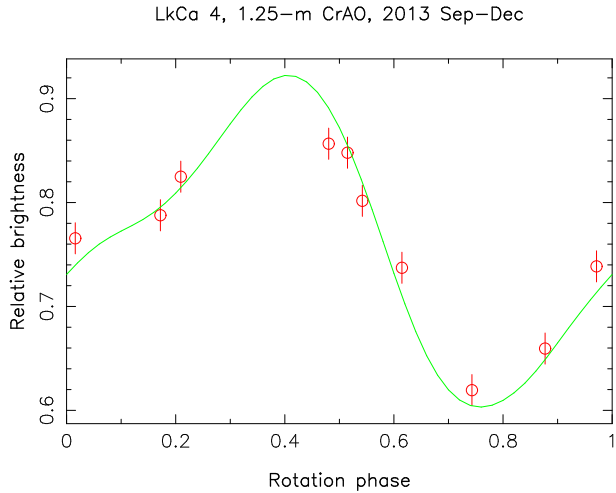}
\caption[]{Brightness variations of LkCa~4 predicted from the tomographic modelling of Fig.~\ref{fig:map} 
of our spectropolarimetric data set (green line), compared with contemporaneous photometric observations in the \V\ band 
(red open circles and error bars) at the 1.25-m CrAO telescope. }
\label{fig:pho}
\end{figure}

We finally stress that reconstructing brightness and magnetic maps simultaneously from both Stokes $I$ and $V$ data (rather than separately from each other 
from either set of profiles) has no more than a small impact on our results;  this is not altogether very surprising since magnetic and brightness imaging are 
largely decorrelated in low-mass stars with large \vsini's, where magnetic and brightness maps depend mostly on Stokes $V$ and Stokes $I$ data respectively 
(rather than more or less equally on both).  More specifically, virtually no change is observed in the brightness map when reconstructing it 
from Stokes $I$ LSD profiles alone, magnetic fields having a much smaller impact than brightness features on disc-integrated line profiles for stars 
featuring high levels of rotational broadening.  The non-simultaneous reconstruction of the brightness map has a larger, though still limited, impact 
on the magnetic map, reducing the strength of both dipole and octupole components (from 1.6 and 0.4~kG down to 1.4 and 0.2~kG respectively) but leaving the overall 
topologies of the poloidal and toroidal fields (and in particular their orientation) unaffected.  We nevertheless think that reconstructing both quantities 
from a simultaneous fit to Stokes $I$ and $V$ LSD profiles provides a more consistent and thus safer approach to the imaging of magnetic stars.  

\subsection{Surface differential rotation}

In addition to showing Stokes $I$ and $V$ signatures detected at a very high confidence level, our spectropolarimetric data set includes observations spread 
over 4 rotation cycles of LkCa~4, making it well suited to look 
for signatures of differential rotation in the way outlined in several previous studies \citep[e.g.,][]{Donati03b, Donati10}.  Practically speaking, 
we achieve this by assuming that the rotation rate at the surface of the star is varying with latitude $\theta$ as  $\omeq - \dom \sin^2 \theta$ 
where \omeq\ is the rotation rate at the equator and \dom\ the difference in rotation rate between the equator and the pole.  The reconstruction code 
uses this law to compute, at each observing epoch, the amount by which each cell is shifted in longitude with respect to the meridian on which 
it is located at the selected reference epoch (chosen at mid-time in our observing run, i.e., cycle 2.0 in the present case). 
% the true location of each elementary region at the surface of the star given its position at the 
% reference epoch at which maps are reconstructed (chosen at mid distance of our observing run, and taken to be cycle 2.0 in the present case). 
This allows the code to properly estimate the spectral contribution of each elementary region at the surface of the star to the synthetic 
disc-integrated Stokes $I$ and $V$ LSD profiles for given values of \omeq\ and \dom.  

For each pair of \omeq\ and \dom\ within a meaningful range, we derive brightness and magnetic maps at a given information content as well as the 
corresponding reduced chi-squared \chisqr\ at which the modelled spectra fit the observations.  We finally obtain \chisqr\ surfaces as a function of 
both \omeq\ and \dom, whose topology is directly indicative of whether differential rotation is detected, e.g., in the case of a convex surface 
featuring a clearly defined minimum.  If so, this \chisqr\ surface can be used to estimate both \omeq\ and \dom\ and their respective error bars by 
determining the location of the minimum and the curvature radii of the \chisqr\ surface at this position.  We usually achieve this by fitting a 
paraboloid to the \chisqr\ surface.  This process has proved reliable for estimating surface differential rotation on various kinds of magnetic 
low-mass stars \citep[e.g.,][]{Donati03b, Donati10}.

The \chisqr\ surface associated with the brightness map features a clear minimum at $\omeq=1.868\pm0.002$~\rpd\ and $\dom=0.010\pm0.006$~\rpd\ (see Fig.~\ref{fig:dr}), whereas 
the magnetic map more or less repeats the same information but with $\simeq$3$\times$ larger error bars ($\omeq=1.865\pm0.005$~\rpd\ and $\dom=0.010\pm0.020$~\rpd).  
These values translate into rotation periods at the equator and pole of $3.364\pm0.004$~d and $3.382\pm0.012$~d respectively, fully consistent 
with the time-dependent periods of photometric fluctuations reported in the literature for LkCa~4 \citep[ranging from 3.367 to 3.387~d,][]{Grankin08} 
known to also probe surface differential rotation.  
This indicates that the surface differential rotation of LkCa~4 is small, much weaker in particular than that of the Sun (equal to $\simeq0.055$~\rpd) 
by typically 5.5$\times$;  in fact, rotation at the surface of LkCa~4 is compatible with solid-body rotation within 1.7$\sigma$, with an 
average rotation period (of $3.369\pm0.004$~d) very close to the one found in the literature value (3.374~d) and used to phase our data (see 
Eq~\ref{eq:eph}).  Brightness and magnetic maps computed assuming our estimate of surface differential rotation are virtually identical to those shown 
previously (see Fig.~\ref{fig:map}).  The small amount of surface differential rotation is likely the reason behind the unusually long lifetime 
of brightness features on LkCa~4 reported above, as was already the case for the mid M dwarf V374~Peg \citep{Morin08a}.  

\begin{figure}
\includegraphics[scale=0.35,angle=-90]{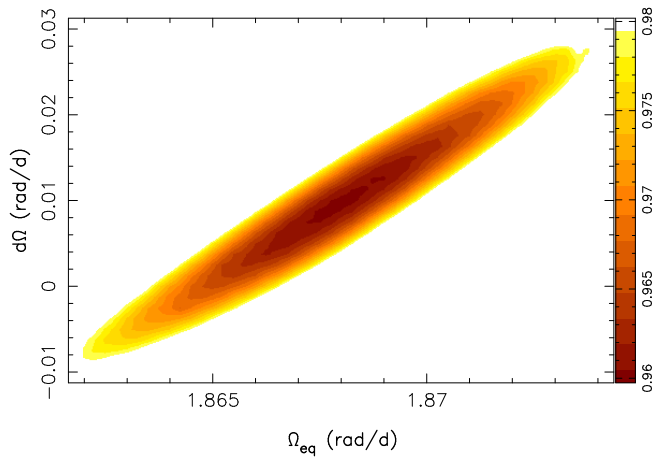}
\caption[]{Variations of \chisqr\ as a function of \omeq\ and \dom, derived from the modeling of our Stokes $I$ LSD profiles at constant information content.  
A clear and well defined paraboloid is observed, with the outer color contour tracing the 2.1\% increase in \chisqr\ that corresponds to a 3$\sigma$ ellipse 
for both parameters as a pair.  This figure is best viewed in color.}
\label{fig:dr}
\end{figure}

\section{Filtering the activity jitter}
\label{sec:fil}

MaTYSSE also aims at detecting potential hJs orbiting wTTSs to quantitatively assess the likelihood of the disc migration scenario in which giant 
protoplanets form in outer accretion discs then migrate inwards until they fall into the central magnetospheric gaps of cTTSs.  If this is indeed 
the case, one expects to find as many hJs as those found around MS stars, and possibly even much more, e.g., to account for those accreted 
by the central star at a later stage of evolution.  However, detecting hJs around wTTSs is not obvious given their very large activity levels often 
generating RV jitters of a few \kms, even at nIR wavelengths \citep[e.g.,][]{Mahmud11, Crockett12}.  MaTYSSE therefore requires that we implement 
an efficient method for filtering out the RV jitter of wTTSs down to a level at which the RV signatures of hJs can be reliably detected.  

We propose to achieve this RV filtering through the tomographic modelling of surface features from phase-resolved spectropolarimetric data sets such as 
that presented herein.  As demonstrated in Fig.~\ref{fig:rv} in the particular case of LkCa~4, this technique is capable of reproducing activity-induced
RV signals at a rms precision of 0.055~\kms, i.e.\ 78$\times$ smaller than the full amplitude of RV variations and less than twice the intrinsic RV 
precision / stability of ESPaDOnS for narrow-lined low-mass stars (of $\simeq$0.03~\kms).  This RV precision is potentially 
enough to detect the presence of hJs, generating RV signatures with a typical semi-amplitude of up to several 0.1~\kms, provided their orbital periods 
are not equal to the rotation period of the central star (or its harmonics).  The technique we propose to implement in practice is to look at rms 
residuals in RV curves of wTTSs once filtered with the tomographic modelling outlined in Sec.~\ref{sec:mod}.  Whenever these residuals happen to be 
larger than 0.1~\kms, we can suspect the presence of a hJ, and even more so if these residuals are found to vary periodically with time at 
a period other than the rotation period of the star and its aliases.  Further confirmation can be obtained by achieving a second spectropolarimetric 
monitoring of the same star to validate the reality of the residual RV signal as well as its planetary origin.  In the specific case of LkCa~4, the 
RV residuals exhibit a rms dispersion of only 0.055~\kms\ with no clear periodic signal with a semi-amplitude larger than 0.10~\kms (see Fig.~\ref{fig:rv});  
we thus conclude that no evidence is found for the presence of a hJ.  

\begin{figure}
\includegraphics[scale=0.35,angle=-90]{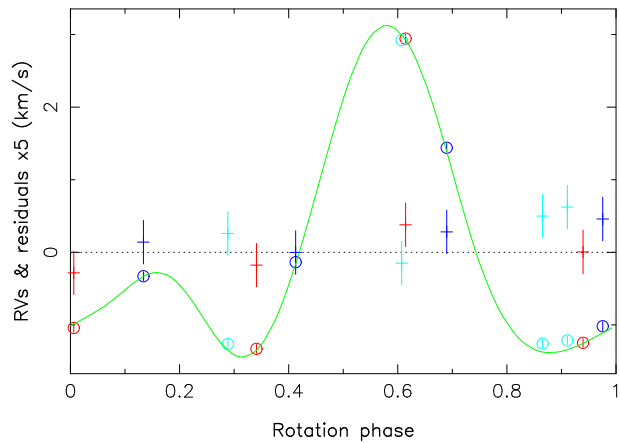}
\caption[]{RV variations of LkCa~4 (in the stellar rest frame) as a function of rotation phase, as measured from our observations 
(open circles) and predicted by the tomographic maps of Fig.~\ref{fig:map} (green line).  RV residuals (expanded by a factor of 5 for 
clarity), are also shown (pluses) and exhibit a rms dispersion equal to 0.055~\kms, i.e., 78$\times$ smaller than the full amplitude of RV variations (of 4.3~\kms).  
Red, dark-blue and light-blue symbols depict measurements secured at rotation cycles 0, 2 and 3 respectively (see Table~\ref{tab:log}).  Given the 
highly distorted shapes of Stokes $I$ LSD profiles (see Fig.~\ref{fig:fit}), RV are estimated as the first order moment of the LSD profile rather than 
through a Gaussian fit to it.  All RV estimates and residuals are depicted with error bars of $\pm$0.06~\kms, estimated from our simulations and reflecting 
the noise level of our data (see text).  This figure is best viewed in color.}
\label{fig:rv}
\end{figure}

Preliminary simulations were carried out to confirm that the method we propose is feasible, and to obtain a first quantitative estimate of its performances.  
In these simulations, we compute a set of synthetic LSD profiles from the image of LkCa~4 we derived (see Fig.~\ref{fig:map}) covering a timespan and featuring 
a noise level similar to that of our observations;  we add to these data a sinusoidal RV signal of semi-amplitude $K$ and period \Porb\ 
simulating the presence of a hJ in close-in circular orbit around the star.  We then reconstruct a brightness image from this data set, use it to filter-out 
the RV curve from the activity jitter and look for periodic signals in the RV residuals.  Results are promising and already confirm that the method we 
propose is valid;  more specifically we find that: 
\begin{itemize}
\item the typical error bar on the RV residuals is 0.060~\kms, i.e., fully compatible with the rms dispersion we measure, and mostly reflects the noise level in the data;  
\item both $K$ and \Porb\ are recovered with reasonable accuracy when $K$ is large enough, when \Porb\ is not too close to the rotation period of the star \Prot\ or 
to its first harmonic (0.5\Prot) and when the phase coverage of the observations is reasonably even;  
\item the 1$\sigma$ error bar on the estimate of $K$ depends both on the number of spectra and on the temporal sampling, decreasing from $\simeq$0.03~\kms\ for a set of $\simeq$16 evenly 
sampled spectra (the standard amount we aim at for MaTYSSE targets) to $\simeq$0.02~\kms\ for twice as much - which should allow us to reliably detect hJs featuring $K$ values of 
0.15 and 0.10~\kms\ respectively;  \item the rms dispersion threshold on the RV residuals (of 0.10~\kms) should be a reliable proxy that a planet is detected.  
\end{itemize}

\begin{figure}
\includegraphics[scale=0.35,angle=-90]{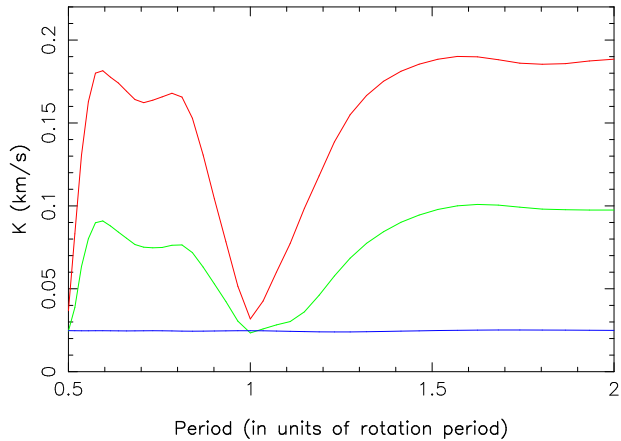}
\caption[]{Semi-amplitude of the recovered RV signal, as a function of the size of the fake RV signal added to our synthetic data (simulating the presence of a hJ, see text) 
and of its period (the orbital period of the simulated hJ, in units of the rotation period of the star), for a simulated data set including 32 spectra evenly sampled over an interval of 
$\simeq$3.5 rotation cycles.  Red, green and blue curves respectively correspond to cases where the added RV signal has a semi-amplitude $K$ of 0.2, 0.1 and 0.0~\kms.  
This figure is best viewed in color.}
\label{fig:sim}
\end{figure}

Of course, these preliminary results (see Fig.~\ref{fig:sim} for an illustration of a typical test) need to be confirmed and validated through a more extensive set of simulations.  
A forthcoming paper will be dedicated to this task;  it will outline in particular the performances of our technique at detecting hJs around wTTSs and at estimating their 
orbital parameters ($K$ and \Porb), and will investigate thoroughly the impact of temporal sampling on these performances.  For now, we can 
already conclude that the tomographic imaging method we described above is efficient at filtering out the activity jitter of wTTSs with extreme levels of 
activity like LkCa~4 and appears as a promising and powerful option for detecting hJs around young Suns.  

For each observing epoch, we are also able to determine the spot-corrected RV of the observed wTTS, i.e., the RV the star would show if no spots nor plages were present 
at its surface.  In the particular case of LkCa~4 in 2014 January, this spot-corrected RV is found to be $16.8\pm0.1$~\kms.  Looking at temporal variations of 
this quantity is yet another way of looking for giant planets orbiting close to very young Sun-like stars.  This method requires multiple data sets such as that
presented here, collected over a time span of a few months, and is thus more time consuming than the approach outlined above;  whenever applicable, it is 
nonetheless very complementary, offering in particular a better sensitivity to massive planets with longer orbital periods (of weeks rather than days).  

\section{Summary \& discussion}
\label{sec:dis}

Our paper presents spectropolarimetric observations of the wTTS LkCa~4 collected with ESPaDOnS at the CFHT in 2014 January, and complemented by 
contemporaneous photometric observations from the 1.25-m telescope at CrAO, in the framework of the MaTYSSE international programme.  

Applying our spectral classification tool to our CFHT data, we first find that LkCa~4 has a photospheric temperature of $4100\pm50$~K and a logarithmic 
gravity (in cgs units) of $3.8\pm0.1$;  this suggests that LkCa~4 is a $0.79\pm0.05$~\msun\ star with a radius of $2.0\pm0.2$~\rsun\ and a rotation 
axis viewed at an inclination angle of $\simeq$70\degr.  With an estimated age of $\simeq$2~Myr, LkCa~4 is still 
fully convective.  Emission in the \caii\ IRT and the \hei\ $D_3$ lines is weak, further confirming the non-accreting wTTS status of LkCa~4.  
In the HR diagram, it sits close to the prototypical cTTSs AA~Tau and BP~Tau, making LkCa~4 a key target for investigating 
how young Suns evolve once they dissipate their accretion discs and start spinning up towards the MS - a crucial transition for our understanding of 
star and planet formation.  
With a rotation period of 3.37~d, i.e., 2.4$\times$ shorter than the typical 8-d rotation period of cTTSs in this range of age and mass, LkCa~4 
is most likely in a process of rapid spin-up.  

Using tomographic imaging on our phase-resolved spectropolarimetric data set, we recovered the brightness map and magnetic topology at the 
surface of LkCa~4 at the epoch of our observations.  
We find that LkCa~4 features a cool spot near the pole as well as a bright plage at intermediate latitudes, covering altogether $\simeq$25\% 
of the overall stellar surface and generating large distortions in the unpolarized LSD profiles at nearly all rotation phases;  the brightness map 
we derive is in good agreement with our contemporaneous photometric observations.  We stress that including both spots and plages in the tomographic 
modeling is necessary to properly fit the observed spectra.  We also find that this brightness distribution experiences 
very little latitudinal shear in the timescale of 4 rotation cycles, implying that differential rotation at the surface of LkCa~4 is $\simeq$5.5$\times$ 
weaker than that of the Sun and compatible with solid-body rotation.  
We obtain that the large-scale magnetic field of LkCa~4 is strong and mainly axisymmetric, including both a $\simeq$2~kG mostly-dipolar poloidal 
field as well as a $\simeq$1~kG toroidal component encircling the star at equatorial latitudes.  The main radial field region is found to overlap
with the cool polar spot in the brightness map; the bright plage is coincident with a region of negative (clockwise) azimuthal field.  

Whereas the nearly-axisymmetric kG poloidal magnetic field is very reminiscent of those of AA~Tau and BP~Tau \citep[e.g.,][]{Donati10b}, the strong 
toroidal field component comes as a major surprise, as no such feature is observed on AA~Tau nor BP~Tau, nor on any of the other fully- and 
mainly-convective cTTSs more massive than 0.5~\msun\ observed to date \citep{Donati13}.  This feature is not observed either on low-mass MS 
stars with similar internal structures \citep{Morin08b}.  We note that the only other (and also fully-convective and fairly young) wTTS for which magnetic 
maps are available in the refereed literature, namely V410~Tau, is also reported to host a significant toroidal field resembling that of LkCa~4 
\citep[][]{Skelly10}, suggesting that what we report here for LkCa~4 is likely a common feature among young fully-convective wTTSs rather than an 
isolated weirdness.  This feature is unlikely to be related to the simple fact that wTTSs rotate significantly faster than cTTSs and to the potential 
differences this faster rotation could generate in the underlying dynamo processes;  dynamos are indeed expected to be saturated in these  
stars and therefore to depend little on the stellar rotation rate.  In fact, no such toroidal field structure is observed in fully-convective M dwarfs, 
even for those with extreme rotation rates \citep{Donati06a, Morin08a}

An alternative option could be that young fully-convective wTTSs like LkCa~4 are triggering non-standard dynamos, as the potential result 
of an unusual internal rotation profile related to the rapid spin-up that wTTSs experience as they contract towards the MS, 
once the braking magnetic torque from the accretion disc no longer operates. 
However, angular momentum redistribution in convective layers is presumably occurring on timescales much shorter than those associated to 
the dissipation of the disc or to the contraction of the star;  
this suggests that wTTSs are unlikely to feature internal rotation profiles that drastically differ from those of MS M dwarfs with 
similar internal structures and rotation rates, and thereby trigger exotic dynamos.  
Our observation that LkCa~4 exhibits no more than a weak surface shear, comparable to that of fully-convective 
M dwarfs \citep[e.g.,][]{Morin08a} and consistent with predictions of recent numerical simulations of fully-convective stars 
\citep[e.g.,][]{Browning08, Gastine12}, brings further support in this direction and provides little help for explaining the surprising 
magnetic topology we unveiled.  Hopefully, this issue will be clarified as MaTYSSE observations pile up and as results for stars of various 
masses, ages and rotation rates are obtained in a consistent way - allowing global trends and new clues to emerge.  

Our tomographic technique is found to be efficient at modelling the activity jitter in RV curves of young active stars, at a rms RV 
precision of 0.055~\kms\ in the particular case of LkCa~4, i.e., 78$\times$ smaller than the full amplitude of the RV jitter.  This performance 
opens promising options for detecting hJs orbiting wTTSs through their periodic signatures in RV curves filtered out from the activity 
jitter.  Preliminary simulations indicate that this technique is indeed capable of detecting hJs providing the semi-amplitude of their RV 
signature is larger than 0.10--0.15~\kms\ (depending on the number of spectra in the data set) and that the planet orbital period is not 
too close to the rotation period of the star (or its first harmonics);  further simulations are needed to confirm and extend these conclusions, 
and will be presented in a forthcoming companion paper.  We report no such evidence for a hJ in the particular case of LkCa~4;  
applying the same analysis to the whole MaTYSSE sample of about 35~wTTSs should allow us to obtain at least an upper limit on the fraction of 
wTTSs hosting hJs, and therefore to assess on observational grounds the standard mechanism proposed to explain the existence of hJs through disc 
migration at a very early stage of star / planet formation.  

On a broader context and a longer timescale, MaTYSSE studies will be extremely useful as science preparation for SPIRou - the nIR spectropolarimeter 
/ high-precision velocimeter presently in construction for CFHT, for a first light in 2017.  With its much higher sensitivity, SPIRou will be able 
to survey a far larger sample of wTTSs, thus expanding by typically an order of magnitude the pioneering work currently being carried out with MaTYSSE.  
Developing and validating activity-modelling and jitter-filtering techniques in the framework of MaTYSSE, such as those explored and described 
in the present paper, will be a key asset for the community as SPIRou comes on line and initiates its Legacy Survey of low-mass protostars.

\section*{Acknowledgements} This paper is based on observations obtained at the Canada-France-Hawaii Telescope (CFHT), operated by the National Research 
Council of Canada, the Institut National des Sciences de l'Univers of the Centre National de la Recherche Scientifique (INSU/CNRS) of France and the University of Hawaii.  
We thank the CFHT QSO team for the its great work and effort at collecting the high-quality MaTYSSE data presented in this paper.  
MaTYSSE is an international collaborative research programme involving experts from more than 10 different countries 
(France, Canada, Brazil, Taiwan, UK, Russia, Chile, USA, Switzerland, Portugal, China and Italy).  
We acknowledge funding from the LabEx OSUG@2020 that allowed purchasing the ProLine PL230 CCD imaging system installed on the 1.25-m telescope at CrAO.
SGG acknowledges support from the Science \& Technology Facilities Council (STFC) via an Ernest Rutherford Fellowship [ST/J003255/1].  
SHPA acknowledges financial support from CNPq, CAPES and Fapemig.
We finally thank the referee, Dr. J.D.~Landstreet, for his detailed reading of the manuscript and for suggestions to improve its overall clarity. 

\bibliography{lkca4}
\bibliographystyle{mn2e}

\end{document}